\documentclass[useAMS,usenatbib]{mn2e}

\usepackage[psamsfonts]{amssymb}
\usepackage[utf8]{inputenc}
\usepackage[T1]{fontenc}
\usepackage[dvips]{graphicx}
\usepackage{amsmath,alltt}
\usepackage{multirow}
\usepackage{rotating}
\usepackage{lscape}

\newcommand{\msun}{\>{\rm M_{\odot}}}

\newcommand{\kms}{\>{\rm km}\,{\rm s}^{-1}}

\newcommand{\bdm}{\begin{displaymath}}
\newcommand{\edm}{\end{displaymath}}
\newcommand{\beq}{\begin{equation}}
\newcommand{\eeq}{\end{equation}}
\newcommand{\bit}{\begin{itemize}}
\newcommand{\eit}{\end{itemize}}
\newcommand{\ben}{\begin{enumerate}}
\newcommand{\een}{\end{enumerate}}
\newcommand{\bfi}{\begin{figure}[htb]}
\newcommand{\bpfi}{\begin{figure}[p]}

\title[Gas depletion in primordial GCs due to accretion onto BHs]{Gas depletion 
in primordial globular clusters due to accretion onto stellar-mass black holes}
\author[Leigh, B{\"o}ker, Maccarone \& Perets]{Nathan W. C. 
  Leigh$^{1}$\thanks{E-mail:nleigh@rssd.esa.int (NL)}, Torsten B\"oker$^{1}$, Thomas J. Maccarone$^{2}$,
Hagai B. Perets$^{3}$ \\
    \\
$^{1}$European Space Agency, Space Science Department, Keplerlaan 1,
2200 AG Noordwijk, The Netherlands \\
$^{2}$School of Physics and Astronomy, University of Southampton,
Highfield, Southampton, SO17 1BJ, United Kingdom \\
$^{3}$Deloro fellow, Physics Department, Technion - Israel Institute of Technology, 
Haifa, Israel 32000}
\begin{document}

\pagerange{\pageref{firstpage}--\pageref{lastpage}} \pubyear{2010}

\maketitle

\label{firstpage}

\begin{abstract}
We consider the effect of compact stellar remnants on the interstellar medium 
of a massive star cluster following the initial burst of star formation.  
We argue that accretion onto stellar-mass black holes is an effective
mechanism for rapid gas depletion in clusters of all masses, as long as they 
contain progenitor stars more massive than $\gtrsim 50\msun$. This scenario 
appears especially attractive for the progenitor systems of present-day massive
globular clusters which likely had masses above M $\gtrsim 10^7\msun$. In
such clusters, alternative mechanisms such as supernovae and stellar winds cannot
provide a plausible explanation for the sudden removal of the primordial gas reservoir 
that is required to explain their complex chemical enrichment history. 
 
In order to consider different regimes in the rate of gas accretion onto stellar mass black holes, 
we consider both the Bondi-Hoyle approximation as well as Eddington-limited accretion. 
For either model, our results show that the cluster gas can be significantly depleted
within only a few tens of Myrs. In addition, this process will affect the distribution of 
black hole masses and, by extension, may accelerate the dynamical decoupling of the 
black hole population and, ultimately, their dynamical ejection. Moreover, the timescales 
for gas depletion are sufficiently short that the accreting black holes could significantly 
affect the chemistry of subsequent star formation episodes.

The gas depletion times and final mass in black holes are not only sensitive to the 
assumed model for the accretion rate, but also to the initial mass of the most massive 
black hole which, in turn, is determined by the upper mass cut-off of the stellar initial mass 
function. Given that the mass function of ``dark'' remnants is a crucial parameter for their 
dynamical ejection, our results imply that their accretion history can have an 
important bearing on the observed present-day cluster mass-to-light ratio. 
In particular, we show that the expected increase of the upper mass cut-off with decreasing 
metallicity could contribute to the observed anti-correlation between the mass-to-light ratio 
and the metallicity of globular clusters. 
\end{abstract}


\section{Introduction} \label{intro}
Over the last few years, observational evidence has accumulated
that few, if any, globular clusters (GCs) consist of a single 
stellar population with a well-defined age and chemical composition.
Instead, multiple stellar populations have been observed in a number of
massive Galactic GCs \citep[e.g.][]{gratton12}. Photometric evidence for 
multiple populations comes in the form of split sequences in the colour-magnitude 
diagram (CMD) during at least one of the various stellar evolutionary phases, 
i.e. main sequence, sub-giant branch, horizontal branch, etc. \citep[e.g.][]{piotto07}. 
More detailed spectroscopic studies have shown that the different stellar populations 
within the same GC often show differences in the relative abundances of light elements 
involved in the proton-capture process such as C, N, O, F, Na, Mg, Al, or Si 
\citep[e.g.][]{osborn71,kraft79, gratton04}.  

This can only be explained if the different stellar populations are born from gas with 
different chemical compositions. A popular scenario to explain the time-variable chemistry 
of the gas reservoir invokes the ``pollution'' of the pristine gas by the chemically 
enriched winds from asymptotic giant branch (AGB) 
stars of the first stellar generation \citep[e.g.][]{gratton01, ramirez02, carretta10}.

This ``self-enrichment'' scenario, however, is complicated by the fact that the spectroscopic
data also demonstrate that the different stellar populations in a GC often show
the same abundances of iron peak and other elements that are produced 
in type II supernovae (SNe II).  The chemical peculiarities also cannot be limited to 
surface contamination, or they would be diluted after the first dredge-up 
episode \citep[e.g.][]{gratton12}.  Currently, no scenario offered to explain the 
presence of multiple populations in GCs is able to fully explain the peculiar abundance 
anomalies, at least not without resorting to a non-canonical initial mass function
\citep[e.g.][]{bekki11,gratton12}.  For
instance, the merging of primordial clusters or gas clumps struggles to
simultaneously explain both the similarities in the $\alpha$-element abundances
and the differences in the light element abundances \citep[e.g.][]{bekki12}.  The 
simplest solution to explain these peculiar abundance patterns is that an 
efficient mechanism for gas depletion must have operated after the formation of the 
first generation, but before the birth of the second generation 
\citep[e.g.][]{conroy11,conroy12}.

Building on previous work \citep{cottrell81,smith87,carretta10}, 
the following picture for the formation of multiple stellar populations in massive GCs 
has therefore been proposed by \citet{conroy11}: the most massive members
of the first stellar generation quickly reach the end of their lifetime and explode as SNe II.  
The energy injected into the ISM from these explosions can remove the material that is 
enriched in $\alpha$-process elements from the GC, and can interrupt the star formation. 
Over time, a second gas reservoir is then created from the ejecta of either AGB stars 
\citep{decressin07} or rotating massive stars \citep{ventura01}, combined with any 
remaining (and/or additional) pristine gas.
After a few hundred Myrs, a second generation of stars is born from this
mix of processed and unprocessed material. To first order, this scenario could 
explain the observed abundance variations in the light elements while
avoiding differences in the $\alpha$-element abundances.

However, the question is whether the SNe II can indeed remove (only)
the gas component that is enriched in $\alpha$-elements.
Theoretical arguments suggest that this is only feasible in clusters less 
massive than $\sim 10^7\msun$ \citep{dopita86,krause12}, because more
massive clusters contain enough gas mass that the SNe II cannot plausibly 
unbind it all. The masses of present-day GCs are 
likely to be only a small fraction of their original mass \citep[e.g.][]{dercole08}.  
It follows that for the most massive Galactic GCs, including $\omega$\,Cen 
\citep[e.g.][]{pancino03}, 
NGC 1851 \citep{bekki12a}, 47 Tuc \citep{milone12a}, NGC 6752 \citep{milone10}, M22 
\citep{milone12b}, and others, 
it appears likely that the initial gas reservoir had a mass of at least
$\sim 10^7\msun$, and probably more. It thus appears doubtful 
that SNe II were able to efficiently remove the $\alpha$-enhanced
gas from these clusters. 

A number of alternative scenarios for the ``purging'' of the gas reservoir in massive
stellar clusters can be found in the literature. For example, \cite{herwig12} invoke Galactic plane 
passages and the associated ram pressure stripping of the gas content to explain the enrichment 
history of $\omega$\,Cen, thus elaborating on earlier work by \cite{tayler75}. 
However, in order for this scenario to explain the multiple stellar populations in an arbitrary GC, 
some fine-tuning between their age difference and the orbital period of the cluster may be required.  
In many cases, the orbital period could be altogether too long for this mechanism to be viable, 
particularly for clusters at large Galacto-centric radii. Another alternative was proposed by \citet{spergel91}
who invokes the relativistic winds of millisecond pulsars as a catalyst for gas expulsion.  Lastly, the feedback 
from x-ray binary jets onto the ISM could also contribute \citep[e.g.][]{fender05,justham12}.

Although the precise mechanism for gas depletion in primordial GCs is currently 
unknown, the effects on the subsequent cluster evolution can be dramatic.  For example, 
\citet{marks08} showed that rapid gas expulsion can unbind a significant fraction of stars 
in the cluster outskirts.  It can even result in complete cluster dissolution, especially 
in the absence of primordial mass segregation \citep[e.g.][]{marks10}.  Thus, in addition 
to accounting for the peculiar abundance anomalies, gas depletion could play a crucial 
role in deciding the present-day relative numbers of first and second generation stars 
in GCs hosting multiple populations \citep[e.g.][]{dercole08,conroy12}.

In this paper, we consider yet another mechanism for the removal
of gas after an initial burst of star formation, namely the
accretion of the ISM onto compact stellar remnants such as black holes
and neutron stars. This scenario is appealing because it can significantly 
deplete the gas reservoir even for very large initial cluster masses, and does not 
require any specific orbital conditions.  
We will show that, in principle, the most massive stellar 
black holes with parent stars at the high end of the initial mass function (IMF)
can be very effective in depleting the gas within the cluster. This can 
significantly modify the final black hole (BH) mass distribution, and therefore 
accelerate the phase of BH-BH ejections that, as theoretical studies suggest, 
should remove most BHs from the cluster \citep[e.g.][]{phinney91,downing10}.

The high-mass end of the stellar initial mass function of primordial GCs and, 
by extension, the number of BH progenitors is highly uncertain  \citep[e.g.][]{maeder09}.  
In open clusters and star-forming associations, however, it has been shown that the maximum 
IMF mass correlates with the total cluster mass, and stars sufficiently massive to be the 
progenitors of BHs should form in clusters more 
massive than $\sim 10^3$ M$_{\odot}$ \citep[e.g.][]{kirk11,kirk12}.  Thus, we 
expect that at least some BHs should 
have formed in a typical primordial GC.  In support of this, there now exists 
compelling observational evidence in favour of BHs residing in 
present-day GCs.  \citet{maccarone07} confirmed the first such BH in 
the giant elliptical galaxy NGC 4472 in the Virgo Cluster.  More recently, 
\citet{strader12} reported two flat-spectrum radio sources in the Galactic globular 
cluster M22, which the authors argue are accreting 
stellar-mass BHs.  This result suggests that the dynamical ejection of BHs from 
clusters may not be as efficient as previously predicted, and that a cluster such 
as M22 could contain on the order of $\sim 5 - 100$ stellar-mass BHs.  Observational 
evidence has also been reported in favour of an intermediate-mass black hole (IMBH) 
existing in the globular cluster M54.  \citet{ibata09} first reported the detection 
of stellar density and kinematic cusps in M54, which they interpreted as being due to 
the presence of a central BH with a mass of $\sim 9400$ M$_{\odot}$.  More recently, 
however, \citet{wrobel11} used Chandra and Hubble Space Telescope astrometry to rule 
out the 
existence of an x-ray counterpart to the proposed IMBH, and placed an improved upper 
limit on its luminosity of L(8.5 GHz) $< 3.6 \times 10^{29}$ erg s$^{-1}$.  Nevertheless, 
the observational evidence supports the presence of at least some BHs in 
present-day GCs. Assuming that BHs are efficiently ejected during the dynamical
evolution of the cluster, it seems likely that a more substantial population of stellar-mass 
BHs once existed in primordial GCs.

Following this introduction, we describe in \S\,\ref{sec:initial} our assumptions for the initial
conditions, i.e. the stellar mass function and dynamical state of the cluster at the time
of BH formation. Using an analytic closed-box model for the evolution of the remnant 
mass function due to accretion from the ISM, we consider two theoretical
models for accretion of ISM onto a black hole, namely the Bondi-Hoyle approximation 
and the Eddington limit. These models, which are described in more detail in \S\,\ref{sec:accmodels},
represent two different mass-dependencies for the accretion rate.  
The results of our analysis, which are presented in \S\,\ref{sec:results}, show that the accretion 
of gas onto compact stellar remnants can have a significant effect on subsequent episodes of 
star formation, and even present-day mass-to-light ratios.  We discuss the implications
of our findings on models of GC formation in \S\,\ref{sec:discussion}, and summarize our
work in \S\,\ref{sec:summary}.

\section{Initial Conditions} \label{sec:initial}
In this section, we present an analytic model for the formation of 
a massive star cluster from its parent giant molecular cloud.
\subsection{Progenitor Mass Function} \label{prog_mass}
We consider a molecular cloud of total mass $M$ that condenses to 
form a star cluster at time $t = 0$.  We assume a star formation 
efficiency (SFE) of 25\%, so that, after the cluster is born, the total mass in stars 
and gas are $M_{\rm s} = 0.25M$ and $M_{\rm g} = 0.75M$, respectively.  
While this value for the SFE agrees with available theoretical constraints 
\citep[e.g.][]{mckee07}, we point out that our results are insensitive to the precise
value of the SFE, as long as it is neither extremely low or extremely high. In the
first limit, the stellar mass of the infant cluster would be too small for many massive 
stars to form, and hence the number of BHs would be limited. At the 
other extreme, an extremely high SFE would leave little gas, and hence the 
BHs could not grow significantly in mass. Our chosen value for the SFE stays
safely away from these regimes.  

We assume an initial mass function for the stellar population according to:
\begin{equation}
\label{eqn:imf}
f_m(m) = \frac{dN}{dm} = {\beta}m^{-\alpha},
\end{equation}
where $N$ is the number of stars with a given stellar mass $m$, and 
$\alpha$ and $\beta$ are constants. We assume a power-law slope of 
$\alpha = 2.3$ \citep{salpeter55}. This choice minimizes the number of 
high-mass stars relative to other IMFs used in the literature, so that 
the estimates we derive in \S\,\ref{sec:results} for the number of stellar 
remnants and the amount of gas accreted by them are conservative. 

As a technical point, the parameter $\beta$ is needed to ensure that the 
correct total stellar mass is preserved when integrating the IMF. It is 
determined by normalizing Equation~\ref{eqn:imf} using the total stellar mass:
\begin{equation}
\label{eqn:imf_norm}
M_{\rm s} = \int_{m_{\rm min}}^{m_{\rm max}} f_m(m)m dm,
\end{equation}
where $m_{\rm min}$ and $m_{\rm max}$ are the minimum and maximum 
stellar masses, respectively.  Integrating Equation~\ref{eqn:imf_norm} 
with respect to $m$ and solving for $\beta$ yields:
\begin{equation}
\label{eqn:beta}
\beta = \frac{M_s(2-\alpha)}{m_{\rm max}^{2-\alpha} - m_{\rm min}^{2-\alpha}}.
\end{equation}
Plausible values for the upper and lower mass cut-off at low metallicity are 
$m_{\rm min} = 0.08$ M$_{\odot}$ and $m_{\rm max} = 150$ M$_{\odot}$ \citep[e.g.][]{dabringhausen11}.  
However, we consider several different values for $m_{\rm max}$ in order to test the 
sensitivity of our results to this assumption.  The mass cut-off values are chosen to be 
conservative for very low metallicity since, as we will show, 
a higher initial maximum BH mass translates into a shorter gas depletion timescale.  It 
follows that our results correspond to upper limits for the gas depletion timescale 
due to accretion onto stellar-mass BHs.

\subsection{Progenitor Velocity Distribution} \label{prog_vel}
In order to calculate the relative velocity between the accretor and the gas 
(which affects the accretion rate) as well as the retention fraction of stellar 
remnants after they experience a potential natal kick, the velocity distribution of
the progenitor population is required. 

In this respect, we assume that the core radius $r_{\rm c}$ is initially comparable to the 
half-mass radius $r_{\rm h}$, since primordial clusters are expected to be more extended 
than their (dynamically evolved) present-day counter-parts \citep[e.g.][]{spitzer87, heggie03}.  
We further assume primordial mass segregation, i.e. that all remnants form inside $r_{\rm c}$, 
and that energy equipartition has been achieved within $r_{\rm c}$ (we will come back to 
these assumptions in \S\,\ref{sec:discussion}).  
%
We therefore adopt a Maxwell-Boltzmann velocity distribution for the progenitors at every 
stellar mass \citep[e.g.][]{binney87}:
\begin{equation}
\label{eqn:MBvel}
f_v(m,v) = N(m)e^{-\frac{1}{2}v^2/\sigma(m)^2},
\end{equation}
where $N(m)$ is the number of stars with mass $m$, $\sigma(m)$ is the 
velocity dispersion for stars of mass $m$, and $v$ is the stellar velocity.  
For the velocity dispersion $\sigma(m)$, we assume
\begin{equation}
\label{eqn:energyequip}
\sigma(m) = \Big( \frac{\bar{m}}{m} \Big) \sigma_0 ,
\end{equation}
where $\sigma_0$ is the central velocity dispersion of the average stellar mass $\bar{m}$.  
For a King model, this is \citep{binney87}:
\begin{equation}
\label{eqn:king_sigma}
\sigma_{\rm 0} = \Big( \frac{4{\pi}G}{9}(\bar{\rho}_{\rm s}+\bar{\rho}_{\rm g})r_{\rm c}^2 \Big)^{1/2},
\end{equation}
where $r_{\rm c}$ is the core radius, $\bar{\rho}_{\rm s}$ is the average stellar
mass density in the core, and $\bar{\rho}_{\rm g}$ is the average gas mass
density in the core.  

\subsection{Progenitor Lifetimes} \label{prog_life}
%
For the progenitor lifetime $\tau_{\rm p}$, we assume that the 
main-sequence (MS) lifetime $\tau_{\rm MS}(m)$ provides a good 
approximation, provided the progenitor mass is $m \le 18$ M$_{\odot}$. This 
seems justified because the MS lifetimes of low-mass stars 
greatly exceed that of every other evolutionary phase, typically
by several orders of magnitude \citep[e.g.][]{clayton68,iben91,maeder09}.  
Note that $\tau_{\rm MS}(m)$ is a function of only the stellar mass.  In 
other words, we are ignoring any metallicity-dependence, since metallicity 
should only weakly affect the total stellar lifetime.  For the MS lifetime, we 
assume \citep{hansen94}:
\begin{equation}
\label{eqn:tau_MS}
\tau_{MS}(m) = \tau_0\Big( \frac{m}{\msun} \Big)^{-2.5},
\end{equation}
with $\tau_{\rm 0} = 10^{10}$ years.  For progenitor masses $m > 18$ M$_{\odot}$, 
we impose a fixed total lifetime of 7 Myrs (note that Equation~\ref{eqn:tau_MS} 
yields the same MS lifetime of 7 Myrs for $m = 18$ M$_{\odot}$).  This is in rough agreement with 
stellar evolution models (Rob Izzard, private communication), which predict a near 
constant lifetime for massive 
stars at low metallicty \citep[e.g.][]{iben91,hurley00,maeder09}.  Thus, our 
final estimate for the total progenitor lifetime is:
\begin{equation}
\label{eqn:age}
\tau_{p} = {\rm max}(\tau_{MS}(m),7\,{\rm Myrs}).
\end{equation}
\subsection{Initial Remnant Mass Function} \label{remnant_mass}
For the initial remnant masses, we use Equations 4,
5, 7, and 9 from \citet{fryer12} for progenitors in the mass
ranges $7.2 \le m$/M$_{\odot} < 11$, $11 \le m$/M$_{\odot} < 30$,
$30 \le$ m/M$_{\odot} < 50$, $50 \le m$/M$_{\odot} < 90$,
and $m \ge 90$ M$_{\odot}$, respectively.  In that order, the initial remnant 
masses are:
\begin{equation}
\label{eqn:fryer1}
M_{r} = 1.35 M_{\odot},
\end{equation}
\begin{equation}
\label{eqn:fryer2}
M_{\rm r} = 1.1 + 0.2e^{(M_{\rm p}-11.0)/4.0} - 2.01e^{0.4(M_{\rm p}-26.0)},
\end{equation}
\begin{equation}
\label{eqn:fryer3}
M_{\rm r} = {\rm min}(33.35 + 4.76(M_{\rm p}-34.0),M_{\rm p} - 0.1(1.3M_{\rm p} - 18.35)), 
\end{equation}
\begin{equation}
\label{eqn:fryer4}
M_{\rm r} = 1.8 + 0.04 \times (90.0 - M_{\rm p}),
\end{equation}
and
\begin{equation}
\label{eqn:fryer5}
M_{\rm r} = 1.8 + \log_{10}(M_{\rm p} - 89.0), 
\end{equation}
where $M_{\rm p}$ and $M_{\rm r}$ are the progenitor and remnant masses, respectively, in units of 
solar masses, and we have used $Z = 0.01Z_{\odot}$, with $Z_{\odot}$ denoting solar 
metallicity. This choice of metallicity roughly agreees with the average for Milky 
Way GCs \citep[e.g.][]{harris96}, and should be appropriate for a star cluster composed 
of Population II stars.  

For each progenitor mass range, two equations are given in \citet{fryer12}
for the initial-final mass relations, one assuming a delayed supernova explosion and the other 
assuming a rapid explosion. The delayed explosion model predicts considerably greater kick
velocities that are roughly consistent with what is observed for pulsars in the Galactic field 
\citep{hobbs05}. Therefore, we assume a delayed supernova explosion in choosing our
initial-final mass relations. 

However, we note that our results are not sensitive to this choice. 
This is because the gas accretion is completely dominated by the most massive BHs, specifically 
those with masses $\gtrsim 50$ M$_{\odot}$. In this high-mass regime, the differences in the 
final remnant mass functions between the delayed and rapid explosion models of \citet{fryer12} 
are negligible. For the same reason, we do not discuss gas accretion onto NSs and white dwarfs (WDs) 
further: they do not affect the results presented here. Nevertheless, in order to check the plausibility
of our assumptions, we compare the retention fractions of NSs to previous estimates in the literature in \S\,\ref{NSs}.


%
\subsection{Remnant Velocity Distribution} \label{remnant_vel}
The accretion rate depends on the relative velocity between the gas 
and the accretor.  Thus, it is important to obtain a 
realistic final remnant velocity distribution from the initial 
progenitor velocity distribution.

The primary factor that determines the initial remnant velocity 
distribution is natal kicks.  These are imparted to most NSs, 
and some (low-mass) BHs, at the time of their formation.  The 
precise physical mechanism responsible is unknown, but appears
to be related to asymmetries in the collapse of the progenitor core, 
or the subsequent supernova explosion \citep[e.g.][]{pfahl02a}.  
Typical kick speeds range from 100 to 200 km s$^{-s}$, although 
radio pulsars with speeds in excess of 1000 km s$^{-1}$ have 
been observed \citep[e.g.][]{lyne94,cordes98,hobbs05}, and kick speeds 
$\lesssim 50 \kms$ are also thought to be possible 
\citep[e.g.][]{blaauw61,pfahl02b,pfahl02c,podsiadlowski04,smits06,paolillo11}.  


We adopt different assumptions for the kick speed for different progenitor 
mass ranges. These assumptions are based on the physical processes 
thought to be driving the kicks.  Here, we present only our assumed 
kick velocities for each mass range, and refer the interested reader to 
\citet{fryer12} for more details.  
Specifically, we assume that no kick occurs for progenitor masses $< 11$ M$_{\odot}$ 
\citep[e.g.][]{podsiadlowski04}.  
%
For progenitors in the mass range $11 \le$ m/M$_{\odot}$ $< 30$ \citep{fryer01}, 
we adopt a kick velocity $v_{\rm kick}$ by randomly sampling from the distribution of observed 
velocities reported by \citet{hobbs05} for radio pulsars in the field of our Galaxy.  
For progenitors with masses $30 \le m$/M$_{\odot}$ $< 40$, we adopt a constant 
kick velocity of $50 \kms$, since the amount of fallback is thought 
to be less sensitive to the explosion mechanism in this range of progenitor 
masses \citep[e.g.][]{fryer12}.  Finally, for progenitors with $m$/M$_{\odot} \ge 40$, 
we assume prompt 
collapse to a BH without a supernova explosion, and thus without any kick.

In order to calculate the initial velocity distribution for our remnant 
population, we add the kick velocity $v_{\rm kick}$ to a randomly drawn 
velocity $v_{\rm 0}$ from Equation~\ref{eqn:MBvel} for the appropriate progenitor 
mass.  This is done using randomized vector addition, since there is no 
reason to expect that the direction of the kick should be correlated with 
the direction of motion of the progenitor through its host cluster.  The 
resulting velocity $v_{\rm rel}$ is then compared to the escape velocities from 
both the core and the cluster.  The latter quantities are calculated using 
the relations
\begin{equation}
\label{eqn:ve-core}
v_{\rm esc,c} = \sqrt{\frac{2GM_{\rm c}}{r_{\rm c}}},
\end{equation}
and
\begin{equation}
\label{eqn:ve-clus}
v_{\rm esc} = \sqrt{\frac{2GM}{r_{\rm c}}},
\end{equation}
where $M_{\rm c}$ is the total (stellar and gas) mass of the core, 
$M$ is the total (stellar and gas) cluster mass, and $r_{\rm c}$ is the 
core radius (recall that we have assumed that $r_{\rm c} \approx r_{\rm h}$ 
initially).  

A comparison between $v_{\rm rel}$ and $v_{\rm esc}$ determines the NS and 
BH retention fractions immediately after their formation.  If 
the final velocity is less than the 
escape velocity of the core (i.e. $v_{\rm rel} < v_{\rm esc,c}$), we 
leave it unchanged from the velocity previously drawn from the 
distribution in Equation~\ref{eqn:MBvel}.  This is because we assume 
that the remnant will rapidly re-achieve energy equipartition with the 
rest of the core, where the local relaxation time is much shorter 
than the global relaxation time \citep[e.g.][]{heggie03}.  If the remnant
ends up with a velocity greater than the escape velocity from the
cluster immediately after it receives its kick (i.e.
$v_{\rm rel} > v_{\rm esc}$), then it is removed completely from the
remnant population.  However, if the post-kick 
velocity exceeds the escape velocity from the core but not that 
from the cluster (i.e. $v_{\rm esc,c} < v_{\rm rel} < v_{\rm esc}$), then 
the remnant will temporarily leave the core 
but still remain bound to the cluster.  In this case, we calculate 
the radius to which the remnant is kicked (called the ``kick 
radius'') using conservation of energy.  This requires two additional 
assumptions, namely that the kick is always imparted at a distance 
$r_{\rm c}/2$ from the cluster centre, and the resulting orbit of the 
kicked remnant within its host cluster is entirely radial.  We then 
calculate the fall-back time $\tau_{\rm FB}$, which decides when the 
remnant will return to the core.  This is given by:
\begin{equation}
\label{eqn:fb}
\tau_{\rm FB} = 2v_{\rm rel}\frac{r_{\rm c}}{GM_{\rm c}}. 
\end{equation}
This assumption is justified given that primordial GCs are thought to have 
been more extended at birth than they are now.  
It follows that stars should be relatively unaffected by dynamical 
friction once outside the core \citep[e.g.][]{mapelli06,leigh11}.  

If $\tau_{\rm MS} + \tau_{\rm FB} < \tau_{\rm f}$, where $\tau_{\rm f}$ 
denotes the time from 
cluster birth until the gas reservoir is depleted, then the remnant 
will return to the core with enough time to accrete at least some 
gas from the dense ISM.  Once in the core, the remnant quickly 
re-achieves energy equipartition with the 
surrounding population.  The remnant's final velocity is then once 
again given by Equation~\ref{eqn:MBvel}.  

\section{Accretion Models} \label{sec:accmodels}

%

The Bondi-Hoyle approximation describes spherically symmetric accretion, and is 
based on the assumption 
that the forces due to gas pressure are insignificant compared to gravitational
forces \citep{bondi44}.  The background gas is treated as
uniform and either stationary or moving with constant velocity
relative to the accretor.  This assumption gives reasonable accretion 
rates, provided the properties of the gas are such 
that the density and total angular momentum are low, at least in the 
vicinity of the accretor.  
%
Numerical studies have confirmed that this is indeed the case, 
provided the gas is not moving rapidly relative to the accretor 
\citep[e.g.][]{fryxell88, ruffert94, ruffert97, foglizzo99}.  
High accretion rates at least approaching the Bondi-Hoyle 
rate have also been observed in several high-mass x-ray binaries 
\citep[e.g.][]{agol02,barnard08}. It thus appears that, at least in 
the low-density, low-angular momentum regime, the Bondi-Hoyle 
approximation is not unrealistic.


The Bondi-Hoyle approximation should be regarded as a strict upper limit to the
true accretion rate\footnote{In this context, it is sometimes useful to distinguish between
a \textit{capture} rate and an \textit{accretion} rate.  The former describes the rate at which
material becomes bound to the accretor, at least temporarily, whereas the latter
describes the rate at which the accretor actually grows in mass.  The Bondi-Hoyle
approximation can provide a good estimate for the capture rate, whereas the
true accretion rate may well be much lower. For the purpose of this paper, however,
we assume that all captured gas is accreted by the BH, which is another argument for
regarding it as an upper limit.}.  
For large accretor masses, the Bondi-Hoyle rate specified in \S\,\ref{sec:accr_rates}
can become extremely high, and pressure forces could play an important role in reducing 
the accretion rate.  
Notwithstanding, theoretical studies have confirmed that super-Eddington accretion is indeed 
possible for massive BHs. For example, although the accretion rates of SMBHs at high redshift are poorly 
constrained \citep[e.g.][]{hopkins10}, some have been observed at redshift $z \approx 6$ 
with masses $\gtrsim 5 \times 10^9$ M$_{\odot}$ \citep[e.g.][]{barth03,willott03}. The presence 
of such massive BHs only $\approx 10^9$ 
years after the Big Bang \citep{king06} is difficult to explain without invoking accretion rates
approaching the Bondi-Hoyle limit. This should be kept in mind when discussing accreting BHs in 
primordial globular clusters, which were also born soon after the Big Bang and at similarly low 
metallicity.  Additionally, photon-trapping can occur at very high accretion rates, which 
makes accretion disks radiatively inefficient and provides a means of circumventing the
Eddington limit \citep{paczynsky80}.  Otherwise, when an accretor is radiating at above the
Eddington luminosity, significant amounts of gas can be expelled at high velocities due to
the intense winds that are initiated \citep[e.g.][]{king03}.

Eddington-limited accretion applies when pressure forces become important, 
and the outward continuum radiation force balances the inward gravitational 
force \citep{eddington26,eddington30}.  The accretion luminosity corresponding to this 
limit is \citep{rybicki79}:
\begin{equation}
\label{eqn:edd}
L_{Edd} = \frac{4{\pi}Gmc}{\kappa},
\end{equation}
where $m$ is the accretor mass, $c$ is the speed of light, and $\kappa$ 
is the electron scattering opacity \citep[e.g.][]{king03}.  
Compared to the Bondi-Hoyle approximation, the Eddington limit provides a much more 
conservative estimate for the accretion rate in the limit of high gas density.  This is 
due the linear dependence on the accretor mass $m$ 
(see \S\,\ref{sec:accr_rates}).  Additionally, the Eddington rate should be much closer to the true 
accretion rate if the gas contains significant angular momentum, and accretion 
proceeds mainly via angular momentum re-distribution within a disk \citep[e.g.][]{king03}.

We assume a constant time-independent background density and
velocity for the accreting gas, since a more sophisticated treatment is
beyond the scope of this paper.  To illustrate the importance of accreting BHs for 
prolonged star formation in primordial GCs, we use the Bondi-Hoyle and Eddington 
approximations to explore two different regimes in the gas 
properties.  The Bondi-Hoyle approximation is representative of the 
gravity-dominated, low-angular momentum regime, whereas the Eddington approximation is 
representative of the pressure-dominated or high-angular momentum regime.

\subsection{Calculating the Accretion Rates}\label{sec:accr_rates}

The Bondi-Hoyle accretion rate is given by Equation 2 of \citet{maccarone12}, which has been re-scaled from 
the formula of \citet{ho03}:
\begin{equation}
\begin{gathered}
\label{eqn:mdot}
\dot{M}_{\rm B-H} = 7 \times 10^{-9} \msun\,{\rm yr}^{-1} \Big( \frac{m}{\msun} \Big)^2 \Big( \frac{n}{\rm 10^6 cm
^{-3}} \Big) \Big( \frac{\sqrt{c_{\rm s}^2 + v^2}}{\rm 10^6 cm s^{-1}} \Big)^{-3} \\
          = A \Big( \frac{m}{\msun} \Big)^2,
\end{gathered}
\end{equation}
where $m$ is the accretor mass, $n$ is the central gas density, 
$c_{\rm s}$ is the sound speed, and $v$ is the relative velocity between
the gas and accretor.  
We assume $n \sim 10^6$ cm$^{-3}$ with central temperatures of a few
thousand K, which translates into $c_{\rm s} \sim 10^6$ cm s$^{-1}$.  These
values were argued by \citet{dercole08} and \citet{maccarone12} to be
representative of the cores of primordial GCs, yielding
$A \approx 10^{-9}$ M$_{\odot}$ yr$^{-1}$ for small accretor masses and 
velocities.  

The Eddington-limited accretion rate, in turn, is given by \citep[e.g.][]{king03}:
\begin{equation}
\label{eqn:edd_mdot}
\dot{M}_{\rm Edd} = \frac{4{\pi}Gm}{{\eta}{\kappa}c},
\end{equation}
where $c$ is the speed of light, ${\eta}c^2$ 
is the accretion yield from unit mass, and ${\kappa}$ is the electron
scattering opacity.  We adopt $\eta = 0.1$ for the accretion efficiency.

%

\subsection{Evolving the Remnant Mass Function} \label{post-accretion-mf}

The most massive stars in the cluster are the first to reach the end of 
their lives due to stellar evolution.  
If gas is still present in significant 
quantities, the most massive stellar remnants (i.e. BHs) should 
begin accreting from the surrounding ISM almost immediately after their formation,
since they do not experience kicks.  

If a kick is imparted, there is a possibly indefinite delay before 
significant accretion begins, because the remnant can be 
expelled from the core.  However, if the remnant remains bound to 
the cluster and the fall-back time $\tau_{\rm FB}$ is sufficiently short, 
then the remnant will eventually begin accreting from the ISM after it 
returns to the central cluster regions.  Thus, accretion is assumed to occur 
only in the core, and to begin after a total time 
$\tau_{\rm i} = \tau_{\rm MS} + \tau_{\rm FB}$, where $\tau_{\rm FB} = 0$ if the 
remnant never leaves the core.  
Once returned to the core, we assume that the remnant remains there.  This is not 
unreasonable since the timescale for accretion to occur 
is comparable to the crossing time of the core, even for accretion rates 
lower than the Eddington rate.  Therefore, the accretor velocity should be 
significantly reduced within a single core crossing time.

The post-accretion mass of a given remnant is calculated as follows.
The total time spent accreting from the ISM can be written as:
\begin{equation}
\label{eqn:mNS}
\tau_{\rm acc} = \int_{\tau_{\rm i}}^{\tau_{\rm f}} dt,
\end{equation}
where the upper limit of integration $\tau_{f}$ corresponds to the time at 
which the post-accretion remnant mass is calculated, and 
$\tau_{\rm i} = \tau_{\rm MS}+\tau_{\rm FB}$.  
Re-writing this in terms of the accretion rate $\dot{m} = dm/dt$ gives:
\begin{equation}
\label{eqn:mNS}
\tau_{\rm acc} = \int_{m_{\rm i}}^{m_{\rm f}} \frac{dm}{\dot{m}},
\end{equation}
where $m_{\rm i}$ and $m_{\rm f}$ are the initial and final remnant masses, respectively.
We then integrate and solve for $m_{\rm f}$ in order to obtain the final 
remnant mass.  Using Equation~\ref{eqn:mdot} for the accretion rate as an 
example, this yields the final remnant mass:
\begin{equation}
\label{eqn:m_final}
m_{\rm f} = \frac{m_{\rm i}}{1 - m_{\rm i}A\tau_{\rm acc}}.
\end{equation}

In our analysis, we have neglected the evolution of the remnant velocity 
distribution during the accretion process even though, in principle, conservation
of momentum causes the BHs to slow down as they accrete mass. 
In the Eddington-limited case, this does not affect the accretion rate, since 
Eq.~\ref{eqn:edd_mdot} is independent of the accretor velocity. In the Bondi-Hoyle 
case, however, a reduction in the accretor velocity {\it increases} the accretion rate, 
and hence the rate of gas depletion.  
Neglecting the BH deceleration, therefore, leads to a lower limit for the
accretion rate in the Bondi-Hoyle approximation, and properly accounting
for these effects would only strengthen our conclusions on the efficiency
of the gas depletion.  

A possibly more important consequence of a lower average BH velocity is that
it could lead to a more efficient migration of massive BHs towards the cluster centre, 
and thus could accelerate the onset of the dynamical BH ejection phase \citep[e.g.][]{phinney91,downing10} 
or the rate of BH-BH mergers near the cluster centre, as discussed in more detail in \S\,\ref{sec:dyn_ej}.
Both effects caused by gas accretion - deceleration and mass growth -  should therefore be taken into 
account when constructing dynamical models of the BH population in gas-rich environments.

\section{Results} \label{sec:results}

\subsection{Black Hole Depletion Times} \label{depletion}

\begin{figure}
\begin{center}
\includegraphics[width=\columnwidth]{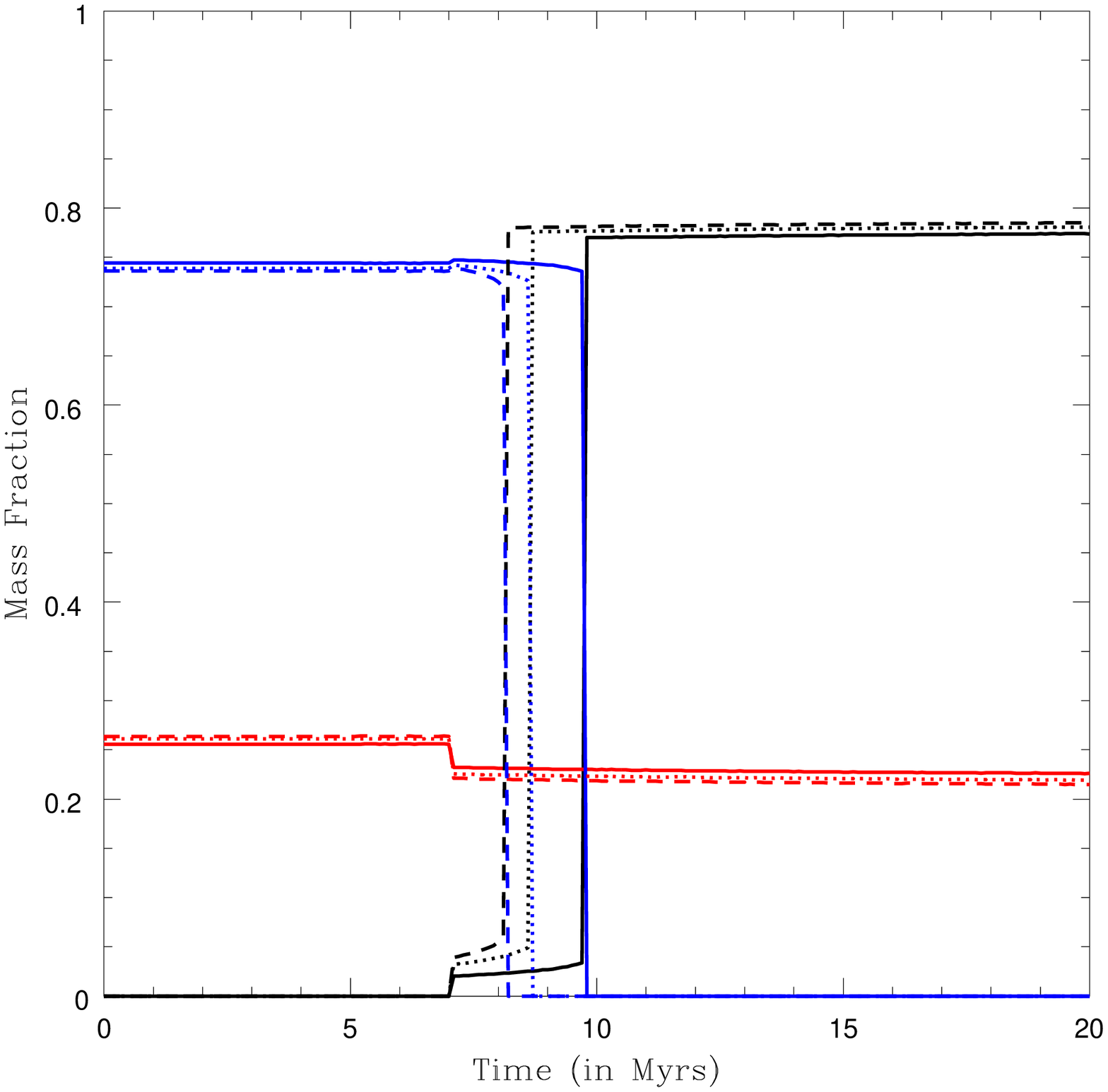}
\end{center}
\vspace*{-0.5cm}
\caption{The relative mass fractions 
in gas (blue), stars (red), and stellar remnants (black) as a function of 
time (in Myrs), calculated using the Bondi-Hoyle approximation to estimate 
the accretion rate. A star formation efficiency of 25\% and an IMF slope of $\alpha = 2.3$ were 
adopted.  The dashed, dotted, and solid lines correspond to an IMF upper 
mass cut-off of 150, 100, and 60 M$_{\odot}$, respectively.  This plot is 
insensitive to our assumption for the total cloud mass $M$, provided 
$M \gtrsim 10^7$ M$_{\odot}$.
\label{fig:massfrac_BH}}
\end{figure}

\begin{figure}
\begin{center}
\includegraphics[width=\columnwidth]{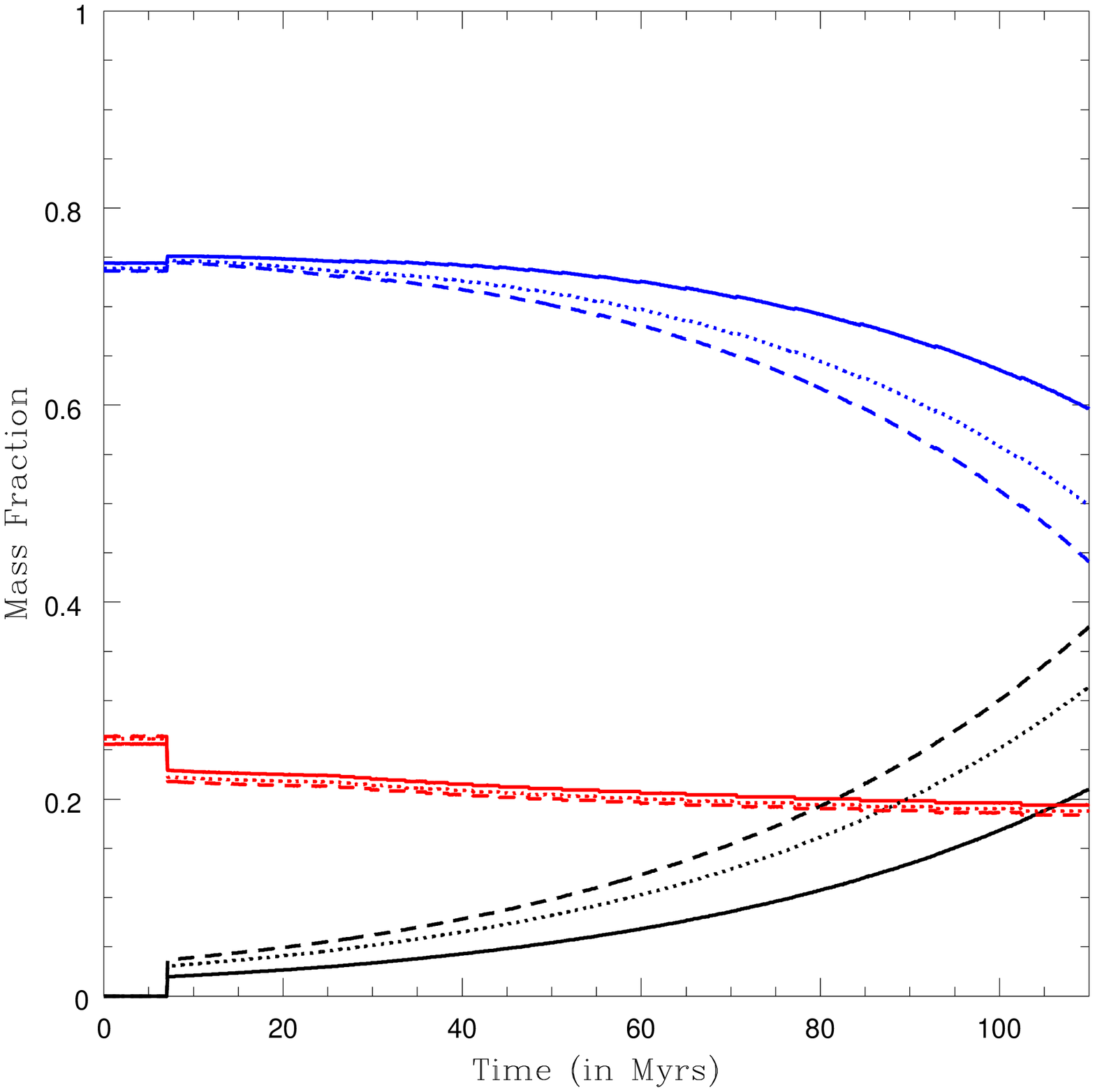}
\end{center}
\vspace*{-0.5cm}
\caption{As Fig.\,\ref{fig:massfrac_BH}, but using the Eddington limit to
estimate the accretion rate.
\label{fig:massfrac_Edd}}
\end{figure}

The results of our analysis are summarized in Figures~\ref{fig:massfrac_BH} 
and ~\ref{fig:massfrac_Edd} which show the time evolution of the relative mass 
fraction of gas, stars, and remnants for Bondi-Hoyle and Eddington-limited accretion, 
respectively.  In both cases, the first BHs form 7 Myrs after the birth of their progenitor 
population, and immediately begin accreting from the ISM.  

In the case of Bondi-Hoyle accretion (Figure~\ref{fig:massfrac_BH}), the steep 
dependence of the accretion rate on black hole mass ($\dot{M}\propto m^2$)
causes the most massive BH(s) to reach very high accretion rates after only a few 
Myrs.  For example, for an IMF upper mass cut-off of 150 M$_{\odot}$ (dashed line), 
the most massive BH reaches an accretion rate of $\dot{M} \approx 10^{-4}$ M$_{\odot}$ yr$^{-1}$ 
after only $\sim 4$ Myrs. At this point, it enters a phase of 
runaway growth and, as shown in Figure~\ref{fig:massfrac_BH}, the remnant mass fraction 
rises asymptotically from only a few percent to nearly 80\% in as little as a few thousand 
years.  
Thus, in the Bondi-Hoyle case, the entire gas reservoir is depleted within a few Myrs 
of cluster formation, nearly independent of our assumption for the IMF upper mass cut-off.  
It follows that the gas depletion time will be under 1 Gyr provided the accretion rate 
$\gtrsim 10^{-2}\dot{M}_{B-H}$.
 
In the case of Eddington-limited accretion (Figure~\ref{fig:massfrac_Edd}), in contrast,
a runaway accretion phase is never reached, because of the weaker (linear) dependence 
of accretion rate on BH mass.  Consequently, the remnant mass fraction rises gradually 
at first over a period of several tens of Myrs, reaching 5-10\% (depending on the assumed 
IMF upper mass cut-off) within $\sim 50$ Myrs of cluster formation.  Nevertheless, the accretion 
rate continues to rise as the BHs grow in mass and, after $\sim 100$ Myrs, the remnant mass 
fraction reaches 20-40\%.  Thus, provided the accretion rate $\gtrsim 10^{-1}\dot{M}_{Edd}$, 
accretion onto stellar-mass BHs will significantly contribute to depleting the available 
gas reservoir within a few hundred Myrs.  

The key conclusion from these plots is that, in both accretion models, stellar-mass BHs 
cause a non-negligible reduction in the total mass of the gas reservoir within as little 
as a few tens of Myrs.  This is the case even if we truncate the IMF upper mass cut-off 
at 60 M$_{\odot}$, which is considerably lower than what is theoretically predicted for 
massive stars at very low metallicity \citep[e.g.][]{iben91,maeder09}.


\subsection{Black Hole Growth} \label{growth}

\begin{figure}
\begin{center}
\includegraphics[width=\columnwidth]{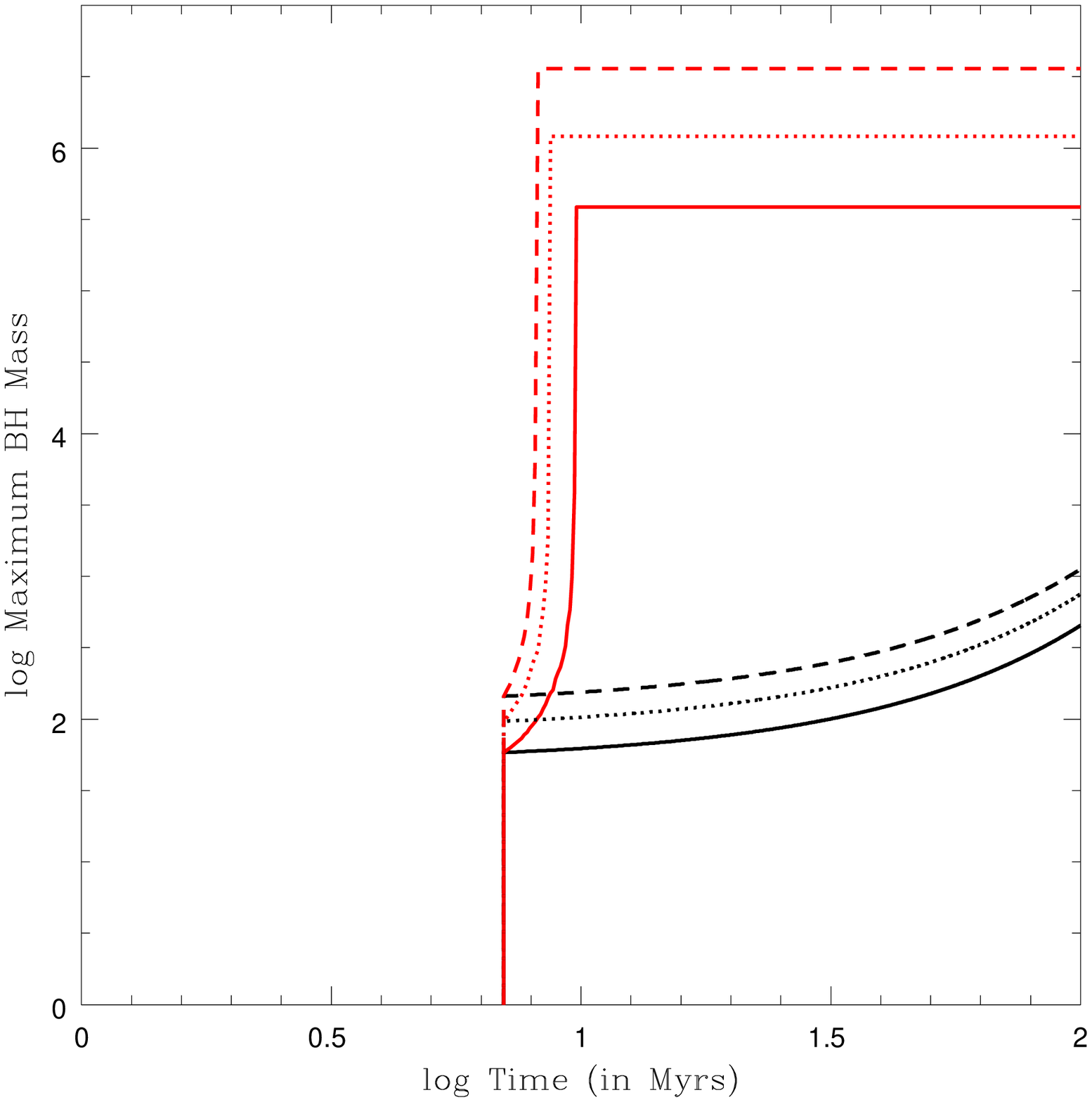}
\end{center}
\vspace*{-0.5cm}
\caption{Mass of the most massive BH as a function of time for the case of
Bondi-Hoyle (red curves) and Eddington-limited (black curves) accretion. 
The dashed, dotted, and solid lines correspond to an IMF upper mass cut-off
of 150, 100, and 60 M$_{\odot}$, respectively.
\label{fig:mmax}}
\end{figure}

The effects of a larger initial BH mass are also demonstrated 
in Figure~\ref{fig:mmax}, 
which shows the maximum BH mass as a function of time for three 
different assumptions for the upper mass cut-off for the IMF, 
namely 150, 100, and 60 M$_{\odot}$.  In the Bondi-Hoyle case, the 
upper limit for each 
mass is obtained by dividing the remaining gas mass by the number of 
BHs in the highest-mass bin (2, 6, and 19 for an IMF upper-mass 
cut-off of 150, 100, and 60 M$_{\odot}$, respectively) 
at the time when runaway accretion sets in.  This represents a 
\textit{theoretical} upper limit to the maximum final BH mass.  We 
caution that, in the Bondi-Hoyle case, such extreme BH masses 
conflict with current \textit{observational} upper limits in 
several massive GCs \citep[e.g.][]{maccarone08,miocchi10,kirsten12,mcnamara12}.  

Our results show that the assumed model for the accretion 
rate and its dependence on accretor mass can have a significant effect
on the final distribution of remnant masses.  
Interestingly, in the Bondi-Hoyle case, the accretion rates can become sufficiently 
high that, if one or a few BHs reach these rates well before their peers, 
there will be a large discrepancy between the final masses of these BHs and the rest.  
It is even feasible that one especially massive BH (or intermediate-mass BH) will 
completely dominate the accretion ``race''. The possible implications of this case
for the subsequent dynamical evolution of the BH population and the evolving 
cluster mass-to-light ratio will be discussed in Section~\ref{sec:discussion}.

Finally, kicks have a minor effect on BH population size due to the fact that 
it is only low-mass BHs that receive kicks.  For initial cloud
masses $M = 10^6, 10^7,$ and $10^8$ M$_{\odot}$, we find BH retention
fractions of roughly 66\%, 90\%, and 96\%  respectively.  Kicks also do 
not significantly affect the gas depletion times since low-mass BHs have 
the lowest accretion rates.  This assumes, however, that the IMF upper mass 
cut-off is sufficiently large that the most massive BHs will not receive 
kicks (see \S\,\ref{remnant_vel}).

\subsection{Neutron Stars} \label{NSs}

Due to their relatively small masses, accretion leaves the numbers of
NSs largely unchanged for up to $\sim 100$ Myrs.  Even in the Bondi-Hoyle 
case, many of the first NSs to form will only accrete enough material within 
this time-frame to end up as especially massive NSs, as opposed to low-mass 
BHs.  Thus, the primary factor affecting NS population size in the core is 
natal kicks.  We can therefore consistently compare 
the NS retention fractions we obtain for our model to previous 
estimates given in the literature calculated using more sophisticated methods.  
This is meant to provide a test of the 
validity of our assumptions.  For initial cloud masses $M = 10^6, 10^7,$ and 
$10^8$ M$_{\odot}$, we find NS retention fractions of roughly
14\%, 37\%, and 73\%, respectively.  
Recently, \citet{ivanova08} used Monte Carlo models for GC evolution to look at NS 
retention due to natal kicks.  Our results are in close agreement with theirs 
for an initial cloud 
mass $M = 10^6$ M$_{\odot}$ with 25\% star formation efficiency, which yields a 
total stellar mass that agrees roughly with the cluster mass considered 
by \citet{ivanova08}.  

\section{Discussion} \label{sec:discussion}

We now discuss the implications of our results 
for the dynamical ejection of BHs, subsequent episodes of 
star formation, and the present-day cluster mass-to-light 
ratio.

\subsection{Dynamical BH Ejection} \label{sec:dyn_ej}

Several theoretical studies suggest that BHs which form soon after
the formation of a massive star cluster should decouple dynamically 
from the rest of the stellar population due to their much 
larger masses.  Consequently, if they don't already form in the cluster
core they will rapidly segregate there \citep[e.g.][]{sigurdsson93,portegieszwart00}.  
Inside the core, BH-BH binaries form dynamically through three-body 
scattering interactions, and most BHs are thought to be eventually 
ejected from the cluster due to strong 
gravitational interactions with other BH-BH binaries.  
However, it is expected to take at least a few Gyrs for
the bulk of the primordial BH population to be depleted in this 
way \citep{oleary06}, and some studies suggest that the required time
could even exceed a Hubble time for some clusters \citep{downing10}.  
Recent studies also suggest that, in general, dynamical BH ejection 
may not be effective enough to remove the entire BH population
from the cluster \citep[e.g.][]{moody09,mackey08}.

In the absence of BH-BH mergers, we expect accretion 
to accelerate the rate of dynamical BH ejections, because mass
growth of a BH will increase its gravitationally-focused cross-section 
for encounters. Consequently, if most BHs experience a significant
mass growth, the average time between close encounters 
(i.e. fly-bys) involving BHs should also decrease. Moreover, 
the velocity of an accreting BH will be reduced significantly 
if its mass increases \textit{on time scales that are short compared
to the central relaxation time} so that it is not re-accelerated by the 
cluster potential.  In this case, conservation of momentum predicts 
that the final post-accretion velocity is:
\begin{equation}
\label{eqn:cons_mom}
v_f = \frac{m_i}{m_f}v_i,
\end{equation}
where $v_{\rm i}$ is the initial BH velocity, and $m_{\rm i}$ and $m_{\rm f}$ 
are the initial (pre-accretion) and final (post-accretion) 
BH masses, respectively.  Such a reduction in the average BH velocity will 
also reduce the average time between BH-BH encounters, which is linearly 
proportional to the relative velocity between interacting objects.
For example, an increase in the average BH mass by a factor of $\sim 10$, 
and the ensuing decrease of the average BH velocity by a factor of $\sim 10$ 
could result in a decrease in the average time between close BH-BH encounters 
by a factor of $\sim 100$. This should probably be regarded as an upper limit, 
however, since mergers between BHs are likely inevitable at very low BH velocities.
Nonetheless, 
we conclude that an episode of accretion from the ISM can significantly reduce 
the BH ejection timescale by a factor of several tens or more.

%

\subsection{Multiple Episodes of Star Formation} \label{sec:mult_sf}

If accreting BHs are indeed responsible for depleting the gas reservoir
of a young stellar cluster, a second generation of stars can only be 
born after most BHs are dynamically ejected, because only then
can gas re-accumulate within the cluster.\footnote{It is sufficient 
that they are kicked to sufficiently large cluster radii that the local 
dynamical friction timescale exceeds the timescale for star formation 
in the core.}  
Previous studies have shown that the amount of gas built up from the winds of 
evolved stars and retained within the cluster can be significant \citep[e.g.][]{naiman12},
and can be sufficient for a second generation of stars to be born.  Moreover, 
in order to explain the peculiar abundance anomalies observed in GCs hosting 
multiple populations, it seems that some pristine gas must also 
somehow re-accumulate within the cluster \textit{after} any polluted 
material due to SNeII from the first generation is removed from the gas 
reservoir \citep[e.g.][]{conroy11,conroy12}.

As discussed in the previous 
section, our results suggest that the epoch of BH ejections 
stimulated via dynamical interactions between BH-BH binaries 
could end much sooner, depending on how much 
the BHs increase their mass via accretion.  Based on the 
results of previous theoretical estimates for this timescale 
\citep[e.g.][]{phinney91,downing10} and typical age spreads estimated 
for multiple populations in massive GCs \citep[e.g.][]{piotto07}, we 
conclude that a scenario in which accreting BHs significantly 
deplete the gas reservoir and are nearly all dynamically-ejected from 
a primordial GC \textit{before} the birth of a second stellar generation 
is not inconsistent with currently available observations and 
theoretical expectations.  

We caution, however, that the timescales for gas depletion and BH ejections 
are somewhat sensitive to the degree of initial mass segregation.  This is because, at 
the large initial cluster masses thought to be needed to explain the observations 
\citep[e.g.][]{conroy11}, the global relaxation time (corresponding to the 
average stellar mass) should be comparable to or exceed the age of the Universe.  
However, whenever a remnant forms, its progenitor is the most massive star in the cluster, 
and many BHs should be born with masses $\sim 100$ times larger than the average 
stellar mass \citep[e.g.][]{fryer12}. Therefore, the progenitors of BHs will likely
migrate into the core first, given that the timescale for mass segregation for a
star of mass $m$ is approximately \citep{vishniac78}:
\begin{equation}
\label{eqn:tau-seg}
\tau_{\rm seg}(m) \approx \frac{\bar{m}}{m}t_{\rm rh}.
\end{equation}
Here, $\bar{m}$ is the average stellar mass and $t_{\rm rh}$ is the half-mass
relaxation time (which can be used to estimate the global rate of two-body
relaxation). Once a star is inside $r_{\rm c}$, the local relaxation time can be
shorter than the global or half-mass relaxation time by a factor of 100 or more  \citep[e.g.][]{spitzer87}.  
The key point is that, with or without primordial mass segregation, the stellar 
remnants of the most massive progenitors are likely to be 
found in the cluster core very early on, and to be dynamically relaxed.

As a mechanism for gas depletion, accretion onto stellar-mass BHs has the 
advantage that it can potentially help to 
explain why stars belonging to the second stellar generations in 
massive GCs do not always show evidence for enrichment in iron-peak 
elements \citep[e.g.][]{gratton12}.  \textit{This is because the accreting BHs 
can remove the polluted material from the SNeII of the first generation 
\textit{before} the second generation is born.}  However, 
we have not considered dynamical ejections that occur 
between massive stars while still on the main-sequence, an effect that
should be considered in future studies.  Finally, the effects of 
accreting BHs on the \textit{first} episode of star formation 
should also be considered, since it may be important if 
star formation is an extended process.  In fact, if the most 
massive stars reach the end of their lives while star formation is still 
ongoing, gas accretion onto their BH remnants could reduce the number 
of first generation stars, and affect the shape of the IMF. 
To summarize, independent of the exact scenario, our results 
demonstrate that accretion onto BHs could significantly affect the star 
formation histories of massive globular clusters.

On the other hand, while the BHs of the 
first stellar generation offer a potentially attractive mechanism for gas depletion, 
their continued gas accretion would pose a problem for the formation 
of the second stellar generation, unless the BHs are efficiently removed 
from the cluster core after their formation. This should be considered in
any proposed scenario explaining the complex star formation history of globular clusters.

\subsection{Black Hole Winds} \label{alt_deplete}
Depleting the gas 
reservoir via accretion does not necessarily require that 
the accreted matter end up bound to the accretor(s).  If
the accretion rate surpasses the Eddington limit, powerful 
winds could develop that might expel 
any remaining gas from the cluster \citep[e.g.][]{king03}.  
It is possible that much of the emission could be in the 
UV, which could in turn result in photoionization 
heating of the ISM beyond the escape velocity from the 
cluster.  Properly quantifying the implications of these 
effects requires detailed numerical 
modeling.  This technology is currently unavailable 
in light of the considerable computational difficulties in 
achieving the necessary resolution \citep[e.g.][]{hopkins10}.

\subsection{The Cluster Mass-to-Light Ratio} \label{mtol}
The considerable uncertainties in the accretion rate onto
BHs, the stellar IMF, and the total amount of accreted 
material make it difficult to accurately predict the final 
mass-to-light ratio of the stellar cluster from our simplistic model.  


In general, if the BHs accrete from the surrounding ISM,
and are retained by their host cluster (e.g. the BHs merge rather 
than being dynamically ejected), one may expect a correlation 
between the cluster metallicity and its mass-to-light ratio.  
This is because theoretical studies suggest that the IMF upper 
mass cut-off increases with decreasing metallicity \citep[e.g.][]{abel02},
which, in turn, accelerates the mass growth of BHs (see \S\,\ref{sec:results}).
Interestingly, such an anti-correlation between the metallicity of globular clusters
and their mass-to-light ratio has indeed been observed by \citet{strader11} in M31.


If, on the other hand, higher accretion rates result in a more efficient dynamical ejection
of BHs, as discussed in \S\,\ref{sec:dyn_ej}, such that (nearly) all BHs are dynamically ejected, 
one may even expect a \textit{correlation} between the cluster metallicity and mass-to-light ratio. 
We conclude that a more sophisticated model is needed to constrain the dependence of the 
final mass-to-light ratio on metallicity.  The key point arising from our model is 
that accretion from the ISM onto a population of remnants in a primordial GC can significantly 
affect its present-day mass-to-light ratio.

\section{Summary} \label{sec:summary}

In this paper, we have considered accreting black holes as a mechanism to 
deplete primordial GCs of their gas after the birth of their progenitor 
population. This mechanism should be effective in clusters of all masses,
as long as they are able to form massive stars that turn into BHs.

We have considered both the 
Bondi-Hoyle approximation and the Eddington limit in order to explore two 
different regimes in the mass-dependency of the accretion rate.  Our results 
suggest that accreting BHs are able to deplete the entire gas reservoir 
in as little as $\sim 10$ Myrs, if the Bondi-Hoyle rate is assumed. 
Even at significantly lower accretion rates on the order of the 
Eddington limit, BHs can accrete a significant 
fraction of the total gas reservoir within a few tens of Myrs. This timescale 
is sufficiently short 
that accreting BHs would significantly impact the chemistry 
of any subsequent episodes of star formation. The accretion rates are 
sensitive to the IMF upper mass cut-off, in the sense that reducing 
the mass of the most massive BH by a factor of $\sim 2$ will result in 
gas depletion times that are longer by a factor of $\sim 2$.

At the end of the accretion period, the average BHs will have significantly increased.
This is likely to accelerate the dynamical decoupling of the BH population
evolution from the rest of the cluster system, and thus may result in a more efficient 
ejection of the BH population from the cluster. We have discussed the implications of 
this effect on subsequent episodes of star formation.

Lastly, we have pointed out that accretion onto stellar remnants may significantly
affect the present-day mass-to-light ratio, although we are currently unable to
precisely quantify the magnitude of this effect.

\section*{Acknowledgments}

The authors wish to thank an anonymous referee for several useful suggestions that improved 
our manuscript.  NL would like to thank Philip Hopkins, Evert Glebbeek, Christian Knigge, 
Morten Andersen, Alison Sills, Guido de Marchi, Rob Izzard, and Philipp Podsiadlowski 
for useful discussions.  TJM would like to thank Stephen Justham for useful 
discussions at the Aspen meeting on the Formation and Evolution of Black Holes in 
2010.  TJM is also grateful to Andrew King, who eloquently explained the difference 
between capture rates and accretion rates.  


\bsp

\label{lastpage}

\end{document}